\documentclass[12pt,preprint]{aastex}
\usepackage{url}
\usepackage[usenames]{color}
\usepackage{amsmath}
\newcommand{\logg} {\log g}

\newcommand{\Te} {T_{\rm eff}}

\newcommand{\msun} {$M_\odot$}
\newcommand{\mh} {M_{\rm H}}

\newcommand{\xcz}{X_{\rm cz}}

\begin{document}

\title{A Convective Dredge-Up Model as the Origin of Hydrogen in DBA White Dwarfs}

\author{B. Rolland, P. Bergeron \& G. Fontaine}
\affil{D\'epartement de Physique, Universit\'e de Montr\'eal,
  C.P.~6128, Succ.~Centre-Ville, Montr\'eal, Qu\'ebec H3C 3J7, Canada}
\email{rolland@astro.umontreal.ca, bergeron@astro.umontreal.ca, fontaine@astro.umontreal.ca}

\begin{abstract}
We revisit the problem of the formation of DB white dwarfs, as well as
the origin of hydrogen in DBA stars, using a new set of envelope model
calculations with stratified and mixed hydrogen/helium
compositions. We first describe an approximate model to simulate the
so-called convective dilution process, where a thin, superficial
hydrogen radiative layer is gradually eroded by the underlying and
more massive convective helium envelope, thus transforming a DA white
dwarf into a DB star. We show that this convective dilution process is
able to account for the large increase in the number of DB white
dwarfs below $\Te\sim 20,000$~K, but that the residual hydrogen
abundances expected from this process are still orders of magnitude
lower than those observed in DBA white dwarfs. Scenarios involving the
accretion of hydrogen from the interstellar medium or other external
bodies have often been invoked to explain these overabundances of
hydrogen. In this paper, we describe a new paradigm where hydrogen,
initially diluted within the thick stellar envelope, is still present
and slowly diffusing upward in the deeper layers of a $\Te\sim
20,000$~K white dwarf. When the convective dilution process occurs,
the bottom of the mixed H/He convection zone sinks deep into the star,
resulting in large amounts of hydrogen being dredged-up to the stellar
surface, a phenomenon similar to that invoked in the context of DQ
white dwarfs.
\end{abstract}


\section{INTRODUCTION}

White dwarf stars represent the endpoint of stellar evolution of more
than 97\% of the stars in the Galaxy, and the determination of their
properties thus provides a variety of useful information on the star
formation history in our galaxy, and on stellar evolution in
general. Having exhausted the nuclear fuel at their center, white
dwarfs enter the cooling sequence at effective temperatures around
150,000 K, and gradually cool off at an almost constant radius over
several billion years. Because of their high surface gravity of the
order of $\logg = 8$, chemical separation induced by gravitational
settling is quite efficient, and the leftover from previous
evolutionary phases are expected to rapidly float to the surface,
while heavier elements sink out of sight, thus producing mostly
hydrogen- or helium-dominated atmospheres (\citealt{Schatzman58};
\citealt{PPFM86}). White dwarfs are found in a variety of flavors: The
DA stars, whose spectra are dominated by hydrogen lines, comprise
about 80\% of the white dwarf population, and their atmospheres are
presumably hydrogen-rich. The remaining 20\% are generally referred to
as non-DA stars. These include the DB (DO) stars, whose spectra are
dominated by neutral (ionized) helium lines, the DQ stars that show
molecular carbon Swan bands, the DZ stars with metallic absorption
features, and the DC stars, which are completely featureless
\citep{wesemael93}. Moreover, additional traces of chemical elements
are also found at the surface of many white dwarfs, producing a large
variety of spectral subtypes (e.g., DAB, DAZ, DBA, DBQ, DBZ, DZA,
etc.) depending on the dominant chemical atmospheric
constituent. However, these various spectral types are found only in
some specific range of temperatures, indicating that there exist
several physical mechanisms that compete with gravitational settling
to alter the chemical composition of the outer layers of white dwarfs
as they evolve along the cooling sequence. Such physical mechanisms
include convective mixing, convective dredge-up from the core,
accretion from the interstellar medium or circumstellar material,
radiative acceleration, and stellar winds \citep{FW87}.

Perhaps the most significant signature of this spectral evolution of
white dwarf stars is the transformation of DA stars into DB stars
below $\Te\sim30,000$~K or so. More than 30 years ago, the
ultraviolet-excess Palomar-Green (PG) survey \citep{GSL86} revealed an
almost complete absence of helium-atmosphere white dwarfs between
$\Te\sim45,000$~K, where the coolest DO stars were found, and
$\sim$30,000~K, where the hottest DB white dwarfs appeared in large
number (see also \citealt{Liebert86}). This so-called DB gap suggested
that a fraction of DA stars, about 20\%, must turn into DB white
dwarfs at some point along the cooling sequence. Even though this gap
has now been partially filled by the discovery of very hot DB stars in
the Sloan Digital Sky Survey \citep[SDSS,][]{EisensteinDB2006}, the
number of DB stars in the gap still remains a factor of 2.5 lower than
what is expected from the luminosity function. The physical mechanism
proposed to explain this DB deficiency as well as the sudden increase
in the number of DB stars is the float-up model \citep{FW87}, where
small amounts of hydrogen thoroughly diluted in the envelope of hot
white dwarf progenitors, slowly diffuse to the surface, thus gradually
transforming a helium-dominated atmosphere into a hydrogen-rich
atmosphere by the time the white dwarf reaches the blue edge of the DB
gap near $\Te\sim45,000$~K. At lower effective temperatures,
$\Te\lesssim30,000$~K, the onset of the helium convection zone would
eventually dilute the superficial {\it radiative} hydrogen layer --- a
process referred to as convective dilution --- thus transforming a DA
star into a DB white dwarf, provided that the hydrogen layer is thin
enough ($\log M_{\rm H}/M_\odot\sim -15$).

A second significant signature of the spectral evolution of white
dwarf stars occurs below $\Te\sim12,000$~K or so, where the fraction
of non-DA to DA white dwarfs increases drastically
(\citealt{Greenstein86}; \citealt{Tremblay08}; \citealt{LBL15}). The
most likely explanation for this phenomenon is a process referred to
as convective mixing, where the bottom of the superficial {\it
  convective} hydrogen layer plunges into the star as it cools off,
eventually reaching the deeper and much more massive convective helium
layer if the hydrogen layer is thinner than $\log M_{\rm
  H}/M_{\star}\lesssim -6$ (\citealt{koester76};
\citealt{vauclair77}; \citealt{dantona79}). At this particular point,
hydrogen and helium are convectively mixed into a single mixed H/He
convection zone. Because the mixing temperature is a function of the
mass of the hydrogen layer (for thicker hydrogen layers, the mixing
occurs at lower effective temperatures), the resulting
hydrogen-to-helium abundance ratio upon mixing is also a function of
the thickness of the hydrogen layer.

Both the convective dilution and the convective mixing processes have
been explored quantitatively by \citet{MV91}, and more recently by
\citet[hereafter RBF18]{Rolland18}. In particular, RBF18 showed (see
their Figure 16) that the convective mixing scenario predicts H/He
abundance ratios upon mixing that agree extremely well with those
determined in cool ($\Te<12,000$~K), He-rich DA white dwarfs where
hydrogen (mostly H$\alpha$) is detected. Also shown is that these
He-rich DA stars will rapidly turn into DC white dwarfs once hydrogen
falls below the limit of detectability. RBF18 also explored
quantitatively the outcome of the convective dilution process, by
predicting the H/He abundance ratios as a function of effective
temperature for various thicknesses of the hydrogen layer (see their
Figure 14), {\it assuming the dilution process has already
  occurred}. The results indicate that hydrogen layer masses between
$\log M_{\rm H}/M_\odot=-13$ and $-10$ (according to ML2/$\alpha=2$
models) are required to account for the presence of hydrogen in the
bulk of DBA stars in the range $20,000~{\rm K}\lesssim \Te\lesssim
12,000$~K, analyzed by RBF18. However, white dwarfs with such large
hydrogen layer mass would not mix until they reach temperatures below
$\Te\sim12,000$~K or so. For the convective dilution process to occur,
the hydrogen layer mass must be much smaller, of the order of $\log
M_{\rm H}/M_\odot\lesssim-15$. For such thin hydrogen layers, the H/He
ratios predicted in the temperature range where most DBA stars are
found are at least three orders of magnitude {\it smaller} that the
observed ratios. In other words, there is much more hydrogen in the
outer layers of DBA stars than expected from a simple convective
dilution process. This problem has been known for a long time (see
\citealt{MV91} and references therein).

Even though there is little doubt that the convective dilution process
is responsible for transforming a significant fraction ($\sim$20\%) of
DA stars into DB white dwarfs, the origin of hydrogen in DBA stars is
still a subject of debate. Since the hydrogen abundances measured in
DBA stars appear too high to have a residual origin, external sources
of hydrogen have often been invoked to account for the observed
abundances, either from the interstellar medium or from other bodies
such as comets, disrupted asteroids, small planets,
etc. (\citealt{MV91}, \citealt{Jura03}, \citealt{veras14}, \citealt{Raddi15},
\citealt{fusillo17}). On the other hand,
\citet{bergeron11} noticed that a lot of the DB white dwarfs in their
sample showed no traces of hydrogen, particularly at low effective
temperatures, with very stringent upper limits.  Such cool
hydrogen-deficient DB stars could only have evolved from hotter white
dwarf progenitors that contain negligible amounts of hydrogen in their
outer stellar envelopes. Based on this observation,
\citet{bergeron11}~found it difficult to reconcile the existence of
these ``pure'' DB white dwarfs with a scenario involving any form of
accretion of hydrogen to explain its presence in the photosphere of
DBA stars. The authors thus concluded that the presence of hydrogen in
DBA stars must be residual, a conclusion also reached by
\citet{KK15}. \citet{bergeron11}~also suggested that perhaps the
convective dilution process is not complete, and that hydrogen somehow
floats on top of the photosphere rather than being forcefully mixed by
the helium convection zone, although \citet{KK15} considered this
scenario unlikely since the convection velocities in the mixed H/He
convection zone are many orders of magnitude larger than the diffusion
velocities of hydrogen in helium.

At the same time, some white dwarfs with helium-dominated atmospheres
have such exceptionally large hydrogen abundances --- which are also
correlated with the presence of metals in large quantities --- that
there is absolutely no doubt that in these cases hydrogen must have
been accreted, most likely from water-rich asteroid debris
(\citealt{Farihi13}; \citealt{Raddi15}, \citealt{fusillo17} and
references therein). Such objects include SDSS J124231.07+522626.6, GD
16, PG 1225$-$079, GD 362, and GD 17, also reproduced in Figure 16 of
RBF18, which shows the exceptionally large hydrogen abundances
measured in these stars with respect to other DBA stars and cool,
helium-rich DA white dwarfs.  More importantly in the present context,
\citet{fusillo17} looked at the population of He-atmosphere white
dwarfs from \citet{KK15}, and found that the incidence of hydrogen was
strongly correlated with the presence of metals, suggesting that
perhaps the majority of helium-rich atmosphere white dwarfs containing
hydrogen and metals are likely to have accreted at least some fraction
of their hydrogen content in the form of water or hydrated minerals in
rocky planetesimal. Since hydrogen will always remain in the mixed
H/He convection zone, while the metals will eventually diffuse away
from the photospheric regions, perhaps this accretion mechanism can
explain the presence of hydrogen {\it in all DBA stars}.

The question of the accretion of hydrogen, either from the
interstellar medium or from other external bodies, has been
investigated by RBF18 --- see their Section 4.4 and Figure 15. By
assuming an average accretion rate over the years, RBF18 found that
the required amount of accreted material, with even a moderate
accretion rate, would build a superficial hydrogen layer thick enough
by the time the white dwarf reaches a temperature of
$\Te\sim30,000$~K, that this object --- presumably a DA star --- would
never turn into a helium-atmosphere DB white dwarf. Based on these
arguments, it was concluded that the hydrogen abundances measured in
DBA stars could not be accounted for by any kind of accretion
mechanism onto a pure helium DB star progenitor. Of course, one
obvious solution around this problem is to invoke a scenario where
accretion of hydrogen-rich asteroid debris begins only {\it after} the
DA-to-DB transition has occurred, in which case accretion would
proceed on a white dwarf with an atmosphere already convectively
mixed. Note that so far, this is the only viable scenario to account
for the large hydrogen abundances measured in the bulk of DBA white
dwarfs. But the question remains: even though there is probably no
doubt that hydrogen is accreted in large amounts in some helium-rich
white dwarfs, is accretion responsible for the presence of hydrogen in
the bulk of DBA stars?

Given the importance of this question, and given the fact that RBF18
used a rather crude envelope models in their exploratory calculations,
we first describe in Section \ref{convdil} improved models that allow
us to simulate the convective dilution process in greater detail. Our
simulations of convective dilution and of accretion onto DB white
dwarfs based on these more realistic models are presented in Section
\ref{results}. We find similar conclusions as RBF18 at the qualitative
level. We next present in Section \ref{dredge} an alternative scenario
where hydrogen might be dredged up from the deep envelope
interior. This is based on the general observation that large amounts
of hydrogen (compared to what is needed to form a DA atmosphere,
$\sim$10$^{-15}$ \msun) might still be diluted in the deep envelope,
especially if a typical DBA white dwarf descend from PG1159 objects as
discussed, for instance, in \cite{quirion12}. We further discuss our
results and conclude in Section \ref{conclu}.

\section{The Convective Dilution Process}\label{convdil}

As discussed above, RBF18 showed in their Figure 14 the outcome of the
convective dilution process --- i.e.~the H/He abundance ratios
predicted at the photosphere as a function of effective temperature
for various thicknesses of the hydrogen layer --- by assuming the
dilution process has already occurred. In other words, the convective
dilution process, per say, was not modeled in any way.
\citet{MV91} used a similar approach, and jumped discontinuously from
hydrogen-rich to helium-rich envelope models (see the case with
$\mh=10^{-14}$ \msun\ in their Figure 1) to describe what these
authors call the convective dredge-up process (see their Table 1).

RBF18 explored the convective dilution scenario by using two sets of
envelope model calculations: models with homogeneous H/He abundance
profiles, and chemically stratified models. In the case of stratified
models, the hydrogen layer of a given mass was forced to sit on top of
the helium layer, in a very approximate way, and was thus never
allowed to mix with the underlying helium envelope. This is obviously
not a physically realistic situation, in most cases. A more suitable
and physical approach is thus required to model properly the kind of
stellar envelopes associated with the convective dilution scenario. We
describe a new approach in this section, which allows us to follow the
convective dilution process in greater detail, and in particular to
predict the mixing temperatures more accurately. The reader interested
only in the results may skip the remainder of this section and go
directly to Section \ref{results} where the most relevant results are
presented.

\subsection{Model Envelope Structures}\label{modstruct}

To explore further the convective dilution process, we use the latest
version of the Montr\'{e}al white dwarf model-building code, which
includes the same input physics as the full evolutionary models
described at length in \cite{FBB01}, with updates discussed
in \cite{vangrootel13}. Here we take advantage of a customized
version of the code in its envelope configuration (see
\citealt{brassard94} for a first description) where an arbitrary
abundance profile can be provided as an input to the model in the form
of the mass fractions of hydrogen, helium, carbon, and oxygen as a
function of depth. If possible, the program will then try to compute a
self-consistent numerical solution with these fixed prerequisites. We
use this particular feature of our code to revise the stratified and
homogeneous models of RBF18, which we describe in turn.

\begin{figure}[bp]
\centering
\includegraphics[keepaspectratio=true,width=0.8\columnwidth]{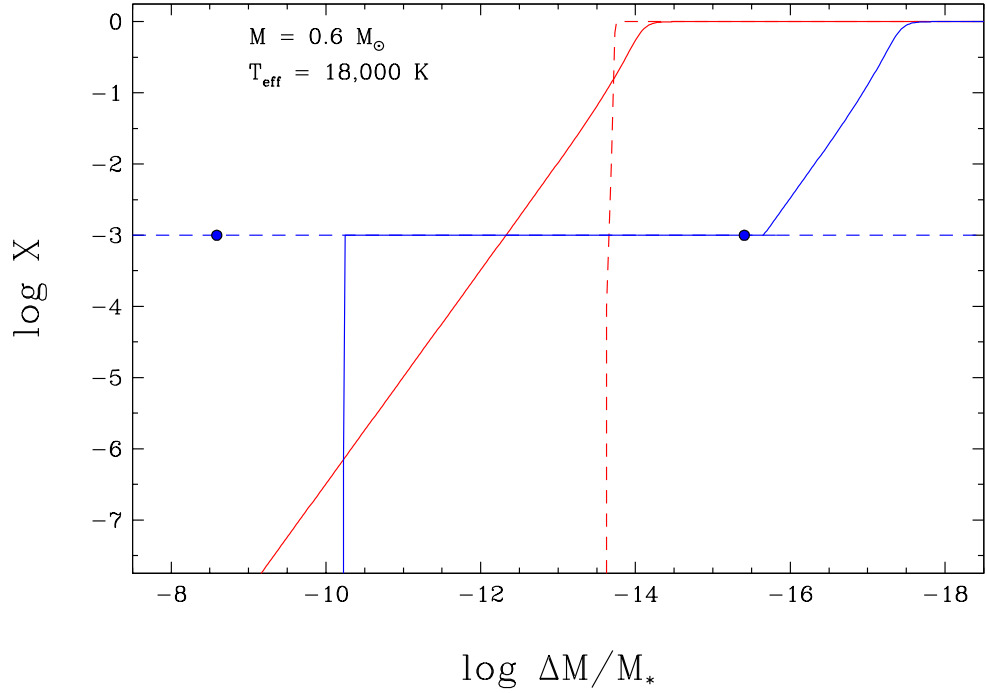}
\caption{Hydrogen mass fraction as a function of depth, expressed as
  the fractional mass above the point of interest with respect to the
  total mass of the star, for various types of envelope structures at
  $\Te=18,000$~K. Chemically stratified ($\log\mh/M_{\star}=-13.77$)
  and mixed H/He models ($\log X=-3$) are shown in red and blue,
  respectively. Our approximate seed models are represented by dashed
  lines, while our improved stratified and hybrid envelope
  structures are shown as solid lines (see text). The extent of
  the convection zone in the seed mixed model is indicated with blue
  dots. All models have been calculated with the ML2/$\alpha=0.6$
  parameterization of the mixing-length theory. \label{com_prof}}
\end{figure}

In RBF18, our stratified models were built using a very rough
approximation, where the hydrogen layer was basically sitting on top
of the helium envelope, leading to spurious convection zones extending
over a few layers near the H/He transition region of our hottest
models (see Figure 11 of RBF18). These where due to too sharp a
transition in composition at the H/He interface. These approximate
thermodynamic structures will be referred to as {\it seed} models for
our improved calculations. Here, we thus go one step further and first
solve the diffusion equation to compute the equilibrium H/He abundance
profile (see, e.g., \citealt{ven88}) using these approximate seed
models. The resulting abundance profile is then fed back into our code
to recompute a new thermodynamic structure that is consistent with
this profile; the entire procedure is repeated until convergence. The
hydrogen mass fraction as a function of depth resulting from this
improved model structure is compared in Figure \ref{com_prof} (red
solid line) with our previous calculations taken from RBF18 (red
dashed line) for a typical 0.6 \msun\ model at $\Te=18,000$~K with
$\log\mh/M_{\star}=-13.77$ ($M_{\star}$ is the mass of the star). As
can be seen, the H/He abundance profile in diffusive equilibrium from
our improved calculations provides a much smoother transition between
the hydrogen and helium layers than our previous models where the
transition region is fairly abrupt.

\begin{figure}[bp]
\centering
\includegraphics[keepaspectratio=true,width=0.8\columnwidth]{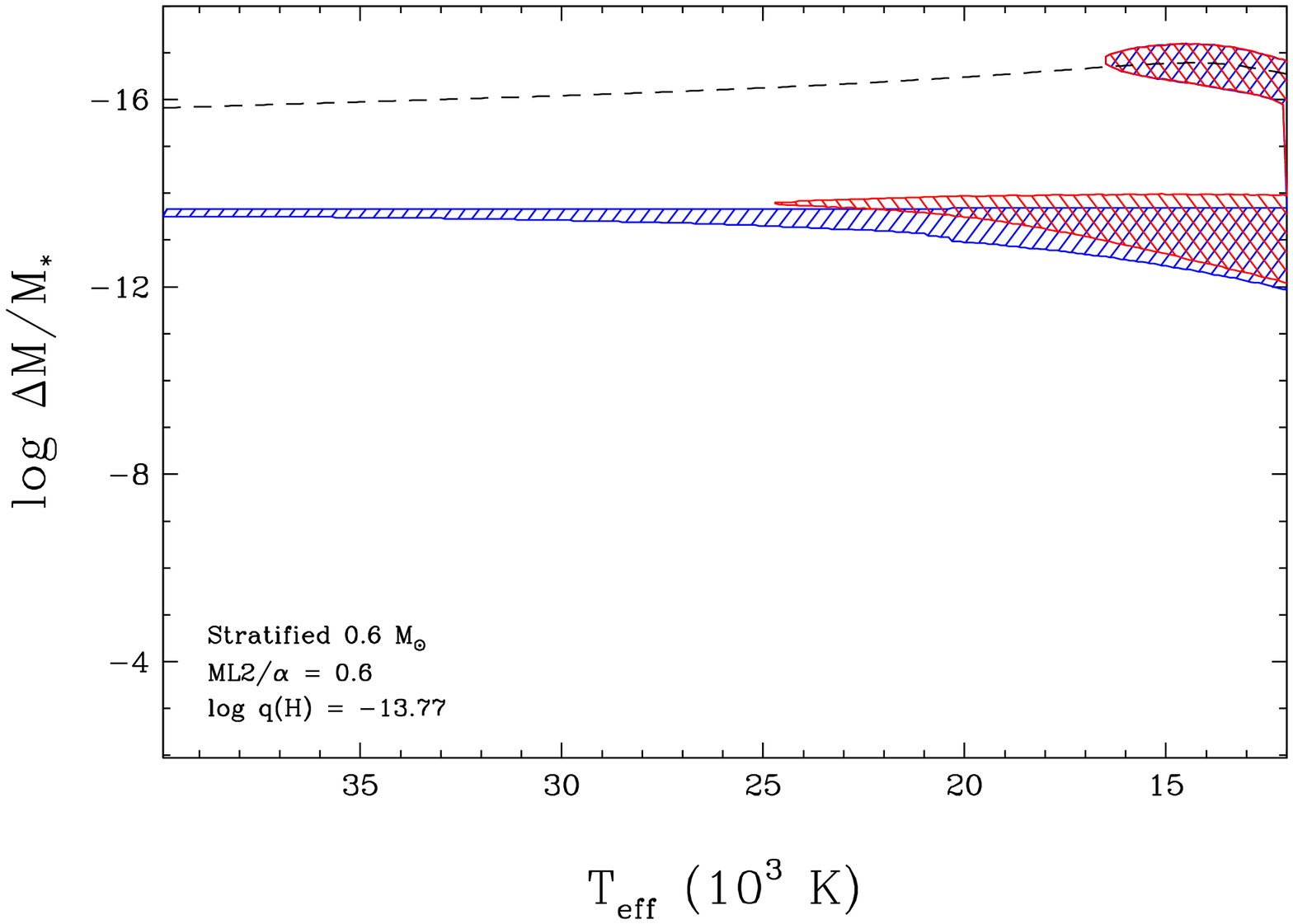}
\caption{Examples of envelope structures for a 0.6 \msun\ white dwarf
  with chemically stratified compositions. The extent of the
  convection zones are shown by the hatched regions. The models shown
  in blue correspond to the original calculations presented in RBF18,
  while those in red are obtained by solving the diffusion equation to
  compute the equilibrium H/He abundance profile (see also Figure
  \ref{com_prof}). The dashed line indicates the location of the
  photosphere in both models.
\label{compar_str_ML2}}
\end{figure}

The effect of these improved calculations on the extent of the
convection zones, as a function of $\Te$, is illustrated in Figure
\ref{compar_str_ML2} for models at 0.6 \msun\ with
$\log\mh/M_{\star}=-13.77$, where our new results are compared with
our earlier calculations presented in RBF18 (see their Figure 11). The
spurious convection zones near the H/He transition region that were
present in our earlier calculations (see the discussion above) have
all but vanished in our improved models. Also, we can see that at
lower effective temperatures, the extent of the helium convective zone
has slightly shifted upwards. Note that in our calculations, we
neglect the fact that the abundance profile should be constant
throughout the convection zone (as opposed to what is displayed by the
red solid line in Figure \ref{com_prof}), because in all cases
explored here, either the extent of the convection zone is extremely
small --- as in Figure \ref{compar_str_ML2} --- or hydrogen remains a
trace element, and is thus not affecting the extent of the convection
zone. For completeness, we would like to point out that the models
displayed in Figure \ref{compar_str_ML2} are comparable to the
solution A shown in Figure 6 of \citet{MV91}, where some small amounts
of hydrogen is mixed within a shallow helium convection zone, but
where most of the hydrogen floats at the surface of the star. 

Note that at some effective temperature and for some thickness of the
hydrogen layer, the deep helium convection zone will become so
efficient as to dilute the superficical hydrogen layer, and the
stratified envelope structures described above will no longer be
appropriate. The fundamental problem is to determine at what
temperature and for which value of the hydrogen layer mass will this
dilution process occur. To answer these questions we develop a new set
of hybrid envelope structures which we now describe.

We thus revisited our chemically homogeneous models from RBF18 (see
their Figures 9 and 10), in which a constant and homogeneous H/He
composition was assumed from the surface to the bottom of the stellar
envelope, characterized by $M_{{\rm env}}/M_\star=10^{-2}$. These
earlier models will now serve as seed models for our improved hybrid
envelope calculations. Here the constant H/He abundance ratio (or
equivalently, the hydrogen mass fraction) throughout the stellar
envelope is replaced by a piece-wise function that follows these
simple criteria: (1) the H/He abundance profile above the convection
zone is assumed to be in diffusive equilibrium\footnote{This
  approximation is identical to that described in \citet{MV91}.}; (2)
the abundance profile must be continuous from the bottom of the
convection zone to the surface; (3) the abundance profile is assumed
to be constant throughout the convection zone, and set to the H/He
abundance ratio of the corresponding seed model; (4) the hydrogen
abundance below the convection zone is set to zero. As discussed in
RBF18, the presence of hydrogen below the convective layer has no
effect on the extent of the convection zone.

An example of this hybrid construct is displayed in Figure
\ref{com_prof} for a 0.6 \msun\ model at $\Te=18,000$~K (blue solid
line); the constant hydrogen mass fraction of the corresponding seed
model is also indicated (blue dashed line). Contrary to our previous
calculations, the hydrogen abundance above the convection zone (the
convection zone is represented by the flat region at $\log X=-3$)
increases steadily towards the surface, as expected from a diffusive
equilibrium profile (see also Figure 6 of \citealt{MV91}). More
importantly in the present context, the presence of hydrogen at the
surface has the effect of modifying the extent of the mixed H/He
convection zone underneath, as indicated by the two blue dots in
Figure \ref{com_prof}, which show the extent of the convection zone in
our homogeneous seed model. In particular, the bottom of the
convection zone has moved significantly higher in the envelope of our
new models. Eventually, if there is too much hydrogen at the surface,
convection in the deeper envelope will become completely
inhibited. When this situation occurs, we should recover the
chemically stratified models described above. This will become the
branching point of the convective dilution process described in
Section \ref{convsim}. Again, we would like to point out that our
improved hydrid model displayed in Figure \ref{com_prof} is
comparable, in this case, to the solution E shown in Figure 6 of
\citet{MV91}, where a large amount of hydrogen is mixed within a deep
helium convection zone, with very little hydrogen floating at the
surface of the star.

Following this procedure, we replaced our original grid of homogeneous
models with a full grid of these hybrid envelope structures assuming
both the ML2/$\alpha=0.6$ and $\alpha=2.0$ parameterizations of the
mixing-length theory to treat convective energy transport. As
discussed in RBF18, these two values bracket the convective
efficiencies mostly used in the context of white dwarf atmospheres and
envelopes. Our independent variable for the construction of these
hybrid models is the hydrogen mass fraction in the convection zone,
labeled $\xcz$\footnote{If more than one convection zone is present,
  the hydrogen mass fraction of the deepest and most massive
  convection zone --- usually associated with He \textsc{ii} --- is
  adopted.}.  Our grid covers a range of effective temperature between
$\Te=12,000~\rm{K}$ and $60,000~\rm{K}$ by steps of $100~\rm{K}$, and
$\log\xcz = -7.75$ to $-0.25$ by steps of $0.25$ dex, thus covering
essentially the entire range between pure helium and nearly pure
hydrogen compositions.

\begin{figure}[bp]
\centering
\includegraphics[keepaspectratio=true,width=0.8\columnwidth]{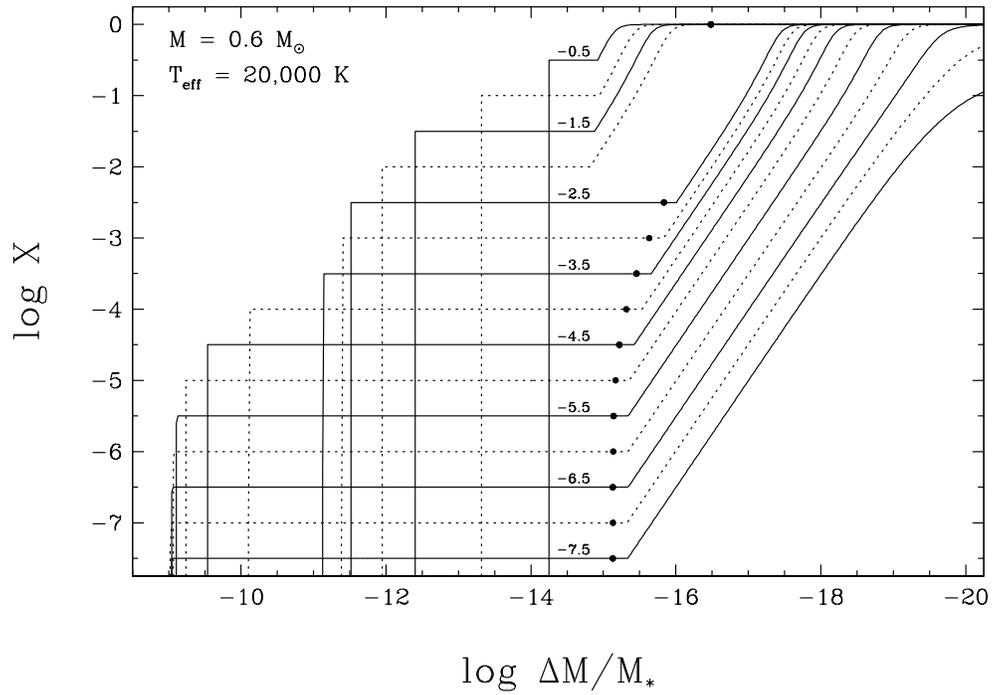}
\caption{Hydrogen mass fraction as a function of depth for our hybrid
  envelope models at 0.6 \msun, $\Te=20,000$~K, and with various
  values of the hydrogen mass fraction in the mixed H/He convection
  zone ($\log\xcz$), labeled in the figure. The black dots indicate
  the location of the photosphere in each model. Only the results for
  ML2/$\alpha=0.6$ are displayed here.
\label{X_prof}}
\end{figure}

The hydrogen mass fraction as a function of depth is displayed in
Figure \ref{X_prof} for 0.6 \msun\ models at $\Te=20,000$~K, and with
various values of the hydrogen mass fraction in the mixed H/He
convection zone ($\xcz$). In each model, the extent of the convection
zone can be identified by the flat part of the abundance profile. Note
that, because of the way these models are constructed, the total
hydrogen mass present in the stellar envelope varies as a function of
effective temperature for a given value of $\xcz$.

\subsection{Total Hydrogen Mass}

\begin{figure}[bp]
\centering
\includegraphics[keepaspectratio=true,width=0.8\columnwidth]{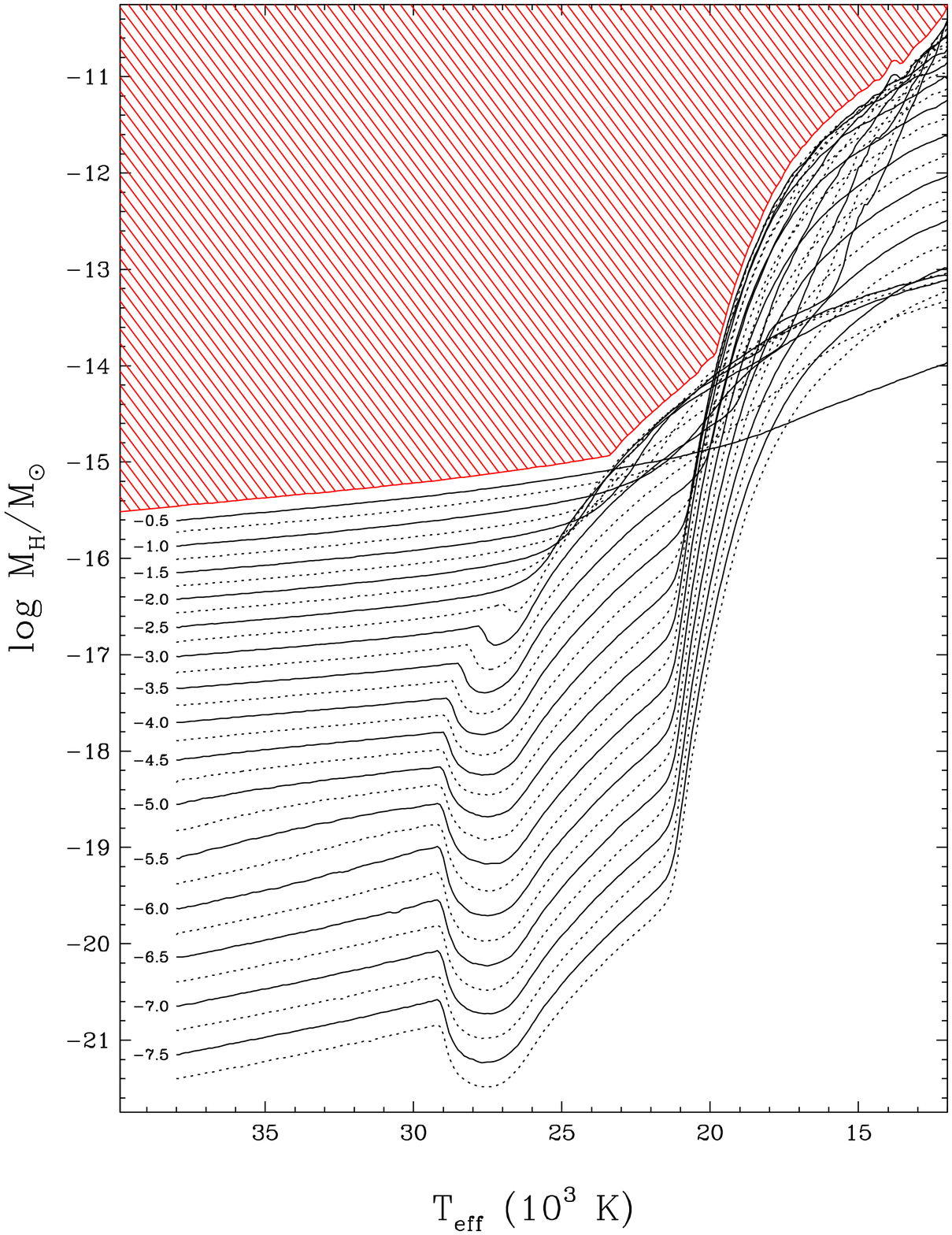}
\caption{Total hydrogen mass, $M_{\rm{H}}$, in our hybrid envelope
  models at 0.6 \msun\ as a function of effective temperature, and for
  various values of the hydrogen mass fraction in the convection zone,
  $\xcz$, labeled on each curve. The region where white dwarfs should
  appear as DA stars is indicated by the red hatched area. Only the
  results for ML2/$\alpha=0.6$ are displayed
  here.\label{mass_cus_ML2}}
\end{figure}

The new calculations described in the previous section allow us to
explore the convective dilution process in greater detail. We remind
the reader that this process describes the outcome of a DA white dwarf
for which the superficial hydrogen layer has been thoroughly diluted
within the underlying helium convection zone. To predict the behavior
of this physical process as a function of time (i.e.,~with decreasing
effective temperature), we must perform a detailed analysis of the
$\Te-M_{\rm{H}}$ parameter space in order to identify the contours of
constant total hydrogen mass. To do so, we first compute the total
amount of hydrogen present in each stellar envelope of our hybrid model
grid. The results of these calculations are presented in Figure
\ref{mass_cus_ML2} where we show the total hydrogen mass content,
$M_{\rm H}$, as a function effective temperature for various values of
the hydrogen mass fraction in the convection zone, $\xcz$. 

The first feature that arises naturally from our calculations is the
existence of an upper limit to the total hydrogen mass that a white
dwarf can contain given our assumed envelope structures. For instance,
there is no envelope structure above $\Te\sim 22,000$~K with a total
hydrogen mass of $\log M_{\rm H}/M_\odot=-14.5$ (for ML2/$\alpha=0.6$
models). Stellar envelopes having such a large, or even larger,
hydrogen mass would require $\log \xcz\rightarrow 0$
(i.e.~$\xcz\rightarrow 1$), and can only exist if their structure is
chemically stratified, corresponding to the DA star configurations
described earlier. Since similar constraints can be obtained at all
effective temperatures, this defines a region in the $\Te-M_{\rm H}$
parameter space, represented by the red hatched area in Figure
\ref{mass_cus_ML2}, where white dwarfs can only exist as chemically
stratified DA stars.

By using the results displayed in Figure \ref{mass_cus_ML2}, it is now
possible to follow the evolution of $\xcz$ as a function of decreasing
effective temperature for a given value of $M_{\rm H}$ by reading the
corresponding value of $\xcz$ at each temperature.  One can see that
there is a unique solution for $\xcz$, at any given effective
temperature, for all values of $\log M_{\rm H}/M_\odot<-16$. For
larger values of $M_{\rm H}$, but below the red hatched region, there
is also a unique solution as long as the temperature remains high
($\Te\gtrsim 25,000$~K).  At lower temperatures and higher total
hydrogen content, however, there are multiple solutions for the same
value of $M_{\rm H}$, generally separated by orders of
magnitude. These degeneracies simply reflect the possibility of mixing
the same total amount of hydrogen in a deep, or in a shallow,
convection zone. Note that \cite{MV91} also found such multiple
solutions for some $\Te$ values --- see in particular their Figures 1
and 6 --- some of which were unstable. In general, there is one
solution corresponding to a DA star configuration (the chemically
stratified model), and one corresponding to a DBA star configuration
(hydrogen appears as a trace in the mixed convection zone), but in
some cases there is another solution where helium is considered a
trace element in diffusive equilibrium within the superficial
hydrogen-rich layer (the solution C in their Figure 6). As discussed
by MacDonald \& Vennes, these last solutions probably do not exist in
nature. In what follows, we use similar considerations to decide which
solution to adopt in our simulations.

\subsection{Convective Dilution Simulations}\label{convsim}

As discussed in the previous sections, if there is too much hydrogen
at the surface of one of our hybrid model structure, convection in the
deeper envelope will become completely inhibited, and at this point we
should recover the chemically stratified models described at the
beginning of Section \ref{modstruct}. Thus, in order to follow the
spectral evolution of white dwarfs with a constant total hydrogen
mass, we start with a chemically stratified model at high effective
temperature $(\Te\sim 60,000~\rm{K})$ and a given value of $M_{\rm
  H}$. We then search a model within our grid of hybrid envelope
structures in the $\Te - M_{\rm{H}}$ parameter space (see Figure
\ref{mass_cus_ML2}) for which the shallow convection zone corresponds
to its stratified counterpart\footnote{Note that such shallow
  convection zones exist even in our chemically stratified structures,
  as shown in Figure \ref{compar_str_ML2}.}. If no such convective
envelope exists, we repeat the procedure with a slightly cooler
stratified model until an appropriate match is found. Since we are
interested in DBA star configurations where hydrogen is always a trace
element, we adopt in Figure \ref{mass_cus_ML2} the appropriate
solution for $\xcz$ that is representative of DBA white dwarfs.  This
approach offers a simple and efficient way to discriminate between the
different $\Te - M_{\rm H}$ degeneracies at a given effective
temperature, and thus ensures a near-continuity in the evolution of
the thermodynamic structures.  From that point on, we only rely on our
hybrid envelope structures. The results of our convective dilution
simulations are presented in the next section.

\section{RESULTS}\label{results}

\begin{figure}[bp]
\centering
\includegraphics[keepaspectratio=true,width=0.8\columnwidth]{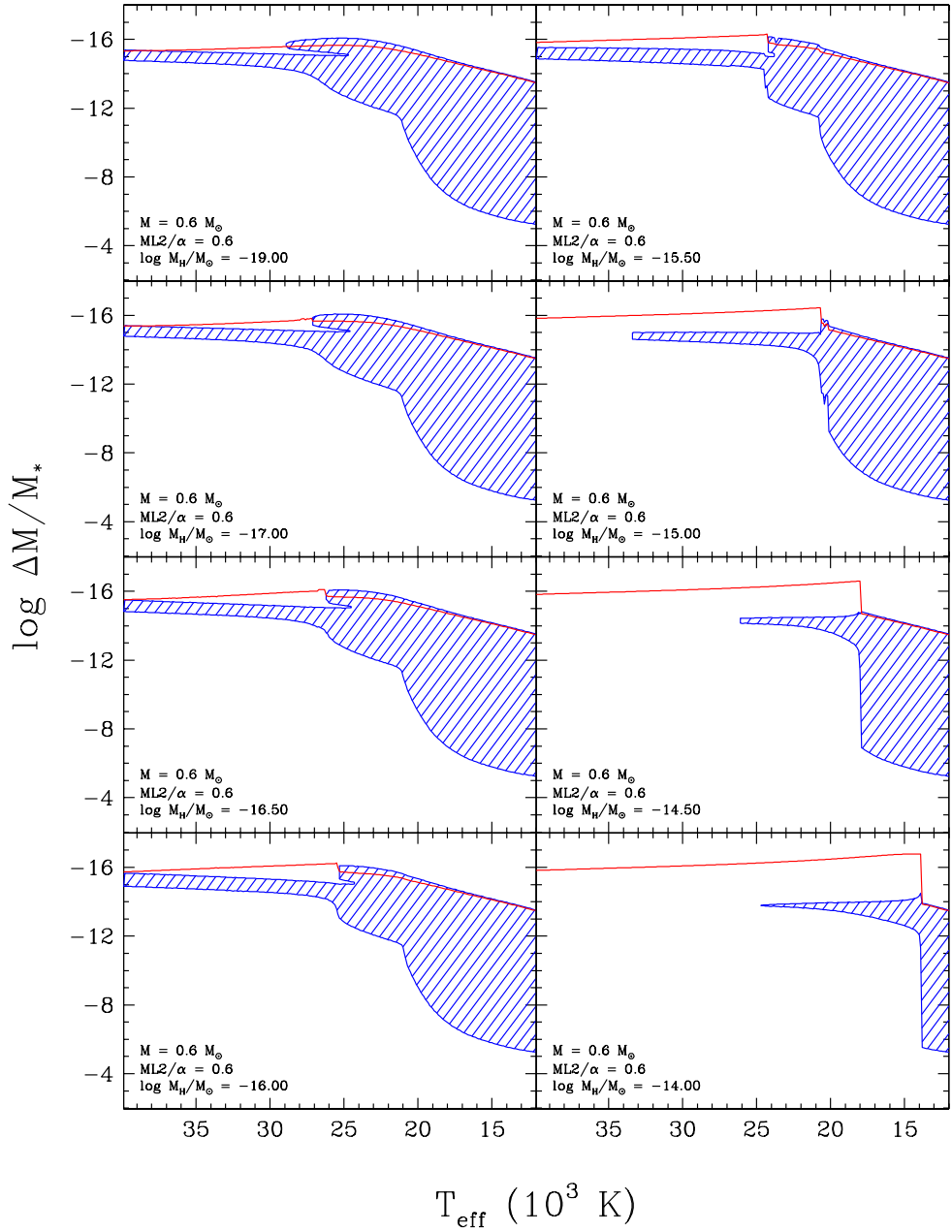}
\caption{Examples of envelope structures for white dwarf models with
  ML2/$\alpha=0.6$ subject to a convective dilution process as a
  function of effective temperature. The depth is expressed as the
  fractional mass above the point of interest with respect to the
  total mass of the star. The red solid line corresponds to the
  location of the photosphere, and the convection zones are shown by
  the hatched region. The models illustrated here are (from upper left
  to bottom right) for 0.6 \msun~with increasing total hydrogen mass
  in the stellar envelope.
\label{envelopes_dil_ML2}}
\end{figure}

\begin{figure}[bp]
\centering
\includegraphics[keepaspectratio=true,width=0.8\columnwidth]{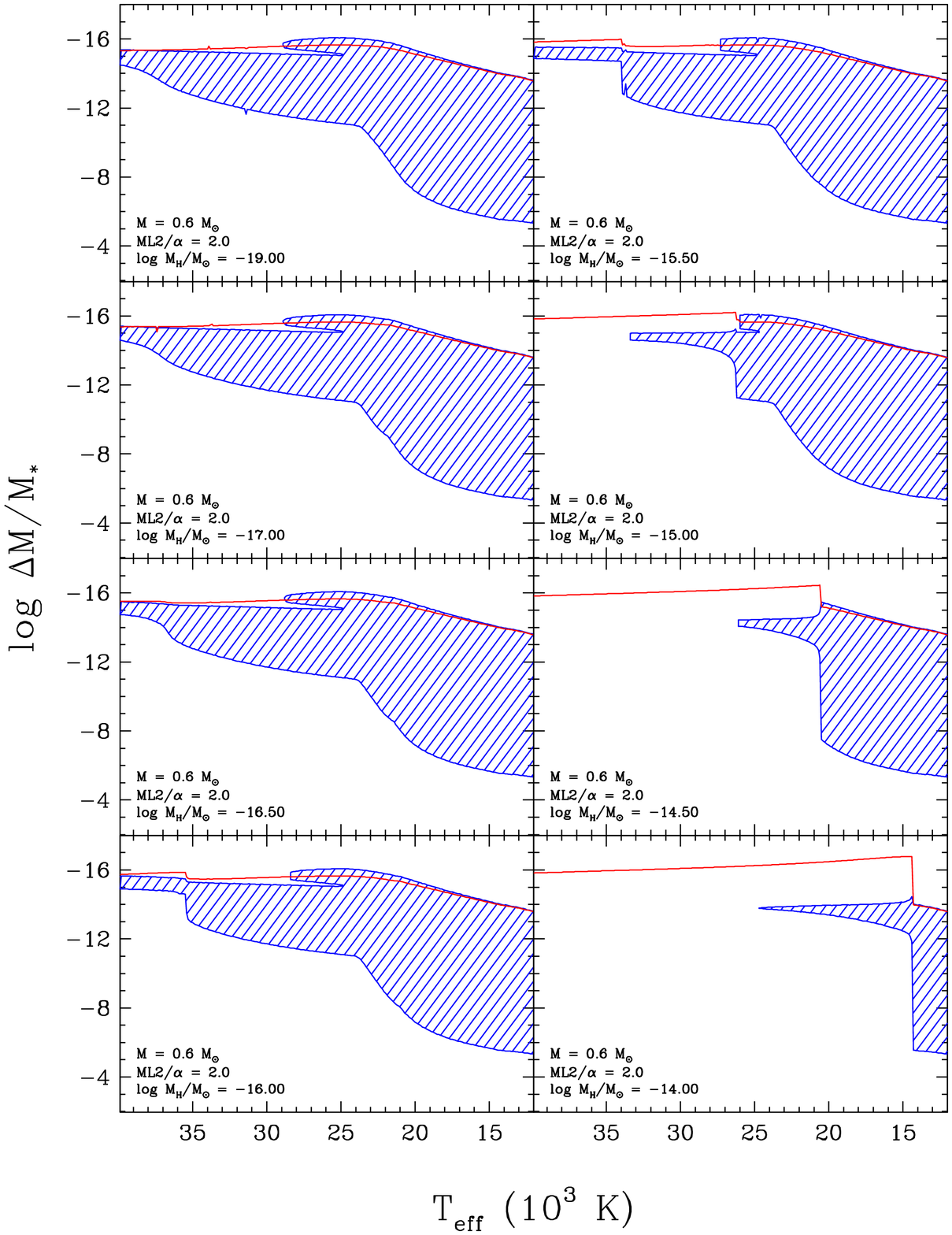}
\caption{Same as Figure \ref{envelopes_dil_ML2} but for ML2/$\alpha=2.0$.
\label{envelopes_dil_ML3}}
\end{figure}

Examples of our envelope structures with constant values of the total
hydrogen mass present in the stellar envelope, $M_{\rm{H}}$, are
displayed in Figures \ref{envelopes_dil_ML2} and
\ref{envelopes_dil_ML3}.  The location of the photosphere as well as
the extent of the convection zones are indicated in each panel. Even
though these sequences represent static envelope models, we believe
they provide a good representation of the convective dilution
scenario, which, in reality, is a much more complex dynamical
process. We first note that, within our framework, white dwarfs with a
total hydrogen mass of $\log M_{\rm H}/M_\odot\geq -13.75$ will never
undergo dilution. Such objects should remain DA stars until the
superficial hydrogen convection zone reaches the deeper convective
helium layers (when $\Te<12,000$~K), at which point convective mixing
occurs if the hydrogen layer is not too thick (see Figure 16 of
RBF18). At the other end of the mass spectrum, our models are almost
identical to pure helium-envelope white dwarfs if $\log
M_{\rm{H}}\lesssim -16.0~M_\odot$. For such thin hydrogen layers,
there is not enough hydrogen accumulated at the surface of the star to
appear as a DA white dwarf (see Figures 3 and 4 of
\citealt{manseau16}). Instead, these objects would appear as
stratified DAB stars, with a thin hydrogen layer floating in diffusive
equilibrium at the stellar surface.

Our version of the convective dilution mechanism should thus be
effective for white dwarf envelopes containing an amount of hydrogen
ranging from $10^{-16}$~\msun\ to $10^{-14}$~\msun. The {\it
  transition temperatures} at which this physical process is expected
to occur are summarized in Table \ref{table_dilT} for both the
ML2/$\alpha=0.6$ and $\alpha=2.0$ parameterizations of the
mixing-length theory.  With ML2/$\alpha=0.6$ for instance, we can see
that depending on the thickness of the hydrogen layer, the DA-to-DB
transition takes place over a narrow range of mixing temperatures,
namely between $\Te\sim25,000$~K and $\sim$14,000 K. The sudden
transition in surface abundance can be observed in Figures
\ref{envelopes_dil_ML2} and \ref{envelopes_dil_ML3} when the
photosphere, indicated by the red line, moves from the radiative
hydrogen layer into the mixed H/He convection zone. As noted by many
authors, including \cite{MV91} and RBF18, there is a direct
correlation between the mixing temperature and $M_{\rm{H}}$, the
latter delaying the growth of the underlying helium convection
zone. Even if our approach is not based on time-dependent
calculations, we can expect the duration of this dynamical
transformation to be much shorter than the typical cooling time,
spanning at most 500 K, or $\sim$10$^{7}$ years. Since the depletion
of the superficial hydrogen layer leads to a rapid growth of the
helium convection zone, convective dilution can be viewed as a
cascade-like process, which becomes increasingly efficient as hydrogen
gradually disappears from the surface.

\begin{deluxetable}{cccccc}
\tablecolumns{4}
\tablewidth{0pt}
\tablecaption{Hydrogen- to Helium-Atmosphere Transition Temperatures\label{table_dilT}}
\tablehead{
\colhead{} &
\multicolumn{1}{c}{$\alpha=0.6$} &
\colhead{} &
\multicolumn{1}{c}{$\alpha=2.0$}\\
\cline{2-2}\cline{4-4}\\
\colhead{log $M_{\rm{H}}$/\msun} &
\colhead{$\Te$ (K)} &
\colhead{} &
\colhead{$\Te$ (K)}
}
\startdata
$-$16.00 & 25,400 & & 35,400 \\
$-$15.75 & 24,800 & & 34,500 \\
$-$15.50 & 24,200 & & 33,900 \\
$-$15.25 & 23,500 & & 27,200 \\
$-$15.00 & 20,600 & & 26,200 \\
$-$14.75 & 19,300 & & 23,600 \\
$-$14.50 & 17,900 & & 20,500 \\
$-$14.25 & 16,100 & & 17,300 \\
$-$14.00 & 13,800 & & 14,300 \\
\enddata
\end{deluxetable}

\begin{figure}[bp]
\centering
\includegraphics[keepaspectratio=true,width=0.8\columnwidth]{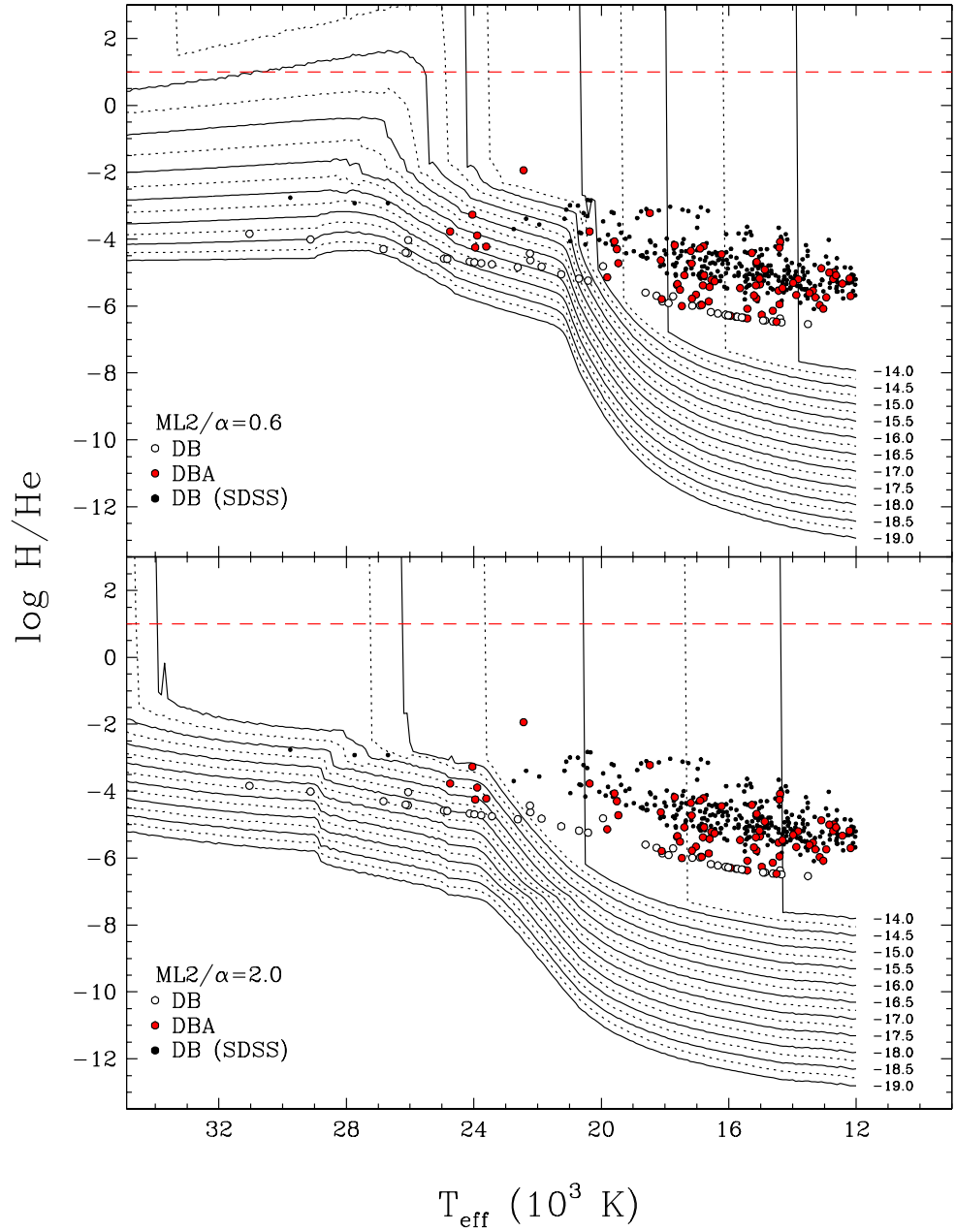}
\caption{Results of our convective dilution simulations for models at
  0.6 \msun~for the ML2/$\alpha=0.6$ and $\alpha=2.0$ versions of the
  mixing-length theory.  Each curve is labeled with the corresponding
  value of $\log M_{\rm{H}}/M_\odot$. The red dashed line indicates
  our empirical limit above which a white dwarf should appear as a DA
  star. Observed hydrogen abundances, or limits, for DB and DBA white
  dwarfs from RBF18 and \cite{KK15} are also
  reproduced.\label{HHe_cus_ML2}}
\end{figure}

We now compare the predictions of our convective dilution model with
the photospheric H/He abundance pattern observed in DB and DBA white
dwarfs. The results of our simulations are presented in Figure
\ref{HHe_cus_ML2}, together with the determination of the
hydrogen-to-helium abundance ratio, as a function of effective
temperature, for all DB and DBA white dwarfs in the sample of RBF18
(see their Figure 5) and in the SDSS sample of \citet{KK15}.  As
discussed above, the sudden drops in the predicted H/He abundance
ratio can be traced back to the results displayed in Figures
\ref{envelopes_dil_ML2} and \ref{envelopes_dil_ML3} when the
photosphere moves from the radiative hydrogen layer into the mixed
H/He convection zone.

Our results indicate that white dwarfs above $\Te\sim 22,000$~K can
easily be explained if the total hydrogen mass is lower than $\log
M_{\rm{H}}/M_\odot\sim -16.5$ with ML2/$\alpha=0.6$ (or lower than
$\log M_{\rm{H}}/M_\odot\sim -15$ with $\alpha=2.0$). As mentioned
above, these objects have never been genuine DA stars at higher
temperatures, and their most likely progenitors are the hot DB stars
identified in the DB-gap by \citet{EisensteinDB2006}. Note that the
few DBA white dwarfs in this temperature range will rapidly evolve
into DB stars, with no detectable traces of hydrogen, as a result of
the growth of the helium convective zone with decreasing effective
temperature, which should efficiently dilute any residual amount of
hydrogen present in the outer layers (see also
\citealt{KK15}). Interestingly enough, the DBA star LP 497-114 (WD
1311$+$129) --- the most hydrogen-rich object in Figure
\ref{HHe_cus_ML2} near $\Te\sim 22,000$~K --- lies close to the
$10^{-15.25}$ \msun\ transformation branch (for ML2/$\alpha=0.6$),
suggesting that this white dwarf might be in the process of being
convectively mixed, as suggested by \citet[][see their Section
  5.4]{bergeron11}, who also reported spectroscopic variations in this
object (see their Figure 28).

The bulk of DB and DBA white dwarfs in Figure \ref{HHe_cus_ML2} is
found below 20,000 K, however, where the bottom of the helium
convection zone sinks rapidly into the stellar envelope. The
significant increase in the number of DB white dwarfs in this
temperature range can be most easily explained by the convective
dilution process, with a total hydrogen mass in the range $-16<\log
M_{\rm H}/M_\odot<-14$ according to our calculations. However, if the
mixing temperatures predicted by our simulations agree well with the
increased number of DB white dwarfs, the amount of residual hydrogen
expected after mixing has occurred is several orders of magnitude
below the observed abundances in DBA stars, a conclusion also reached
in all previous investigations (see RBF18, \citealt{MV91}, and
references therein). As discussed by RBF18, in order to match the
observed hydrogen abundances in DBA white dwarfs, hydrogen layer
masses larger than $\log M_{\rm H}/M_\odot\sim-13$ are required, in
which case the convective dilution process becomes impossible.

We must therefore conclude that most, but not all, helium-atmosphere
white dwarfs cooler than $\Te\sim 30,000~\rm{K}$ containing traces of
hydrogen cannot be explained in terms of our version of the convective
dilution scenario. The most common solution proposed to solve this
problem is to assume that a significant fraction of DB stars are
indeed the result of a convective dilution scenario, with DA
progenitors having very thin hydrogen layers $(\log
M_{\rm{H}}/M_\odot\leq -15)$, but that after the DA-to-DB transition
has occurred, accretion of hydrogen from the interstellar medium or
other external bodies (comets, disrupted asteroids, etc.) increases
the hydrogen content in the stellar envelope, up to the level observed
in DBA white dwarfs. But as discussed in RBF18 (see their Figure 15),
the amount of accreted material required to account for the observed
abundances in DBA stars will most likely build a thick enough hydrogen
layer in the earlier evolutionary phases, such that the convective
dilution process will again become impossible. 

To study the effect of the accretion of hydrogen from external sources
more quantitatively, we repeated our previous calculations by
considering a constant accretion process onto a pure helium-envelope
white dwarf at 0.6 \msun. We considered various accretion rates ranging
from $\Dot{M}=10^{-27}$ \msun\ yr$^{-1}$ to $10^{-16.5}$
\msun\ yr$^{-1}$ by steps of 0.25 dex, and calculated the total
accreted mass of hydrogen as a function of the white dwarf cooling
time, or equivalently, as a function of effective temperature. 

\begin{figure}[bp]
\centering
\includegraphics[keepaspectratio=true,width=0.8\columnwidth]{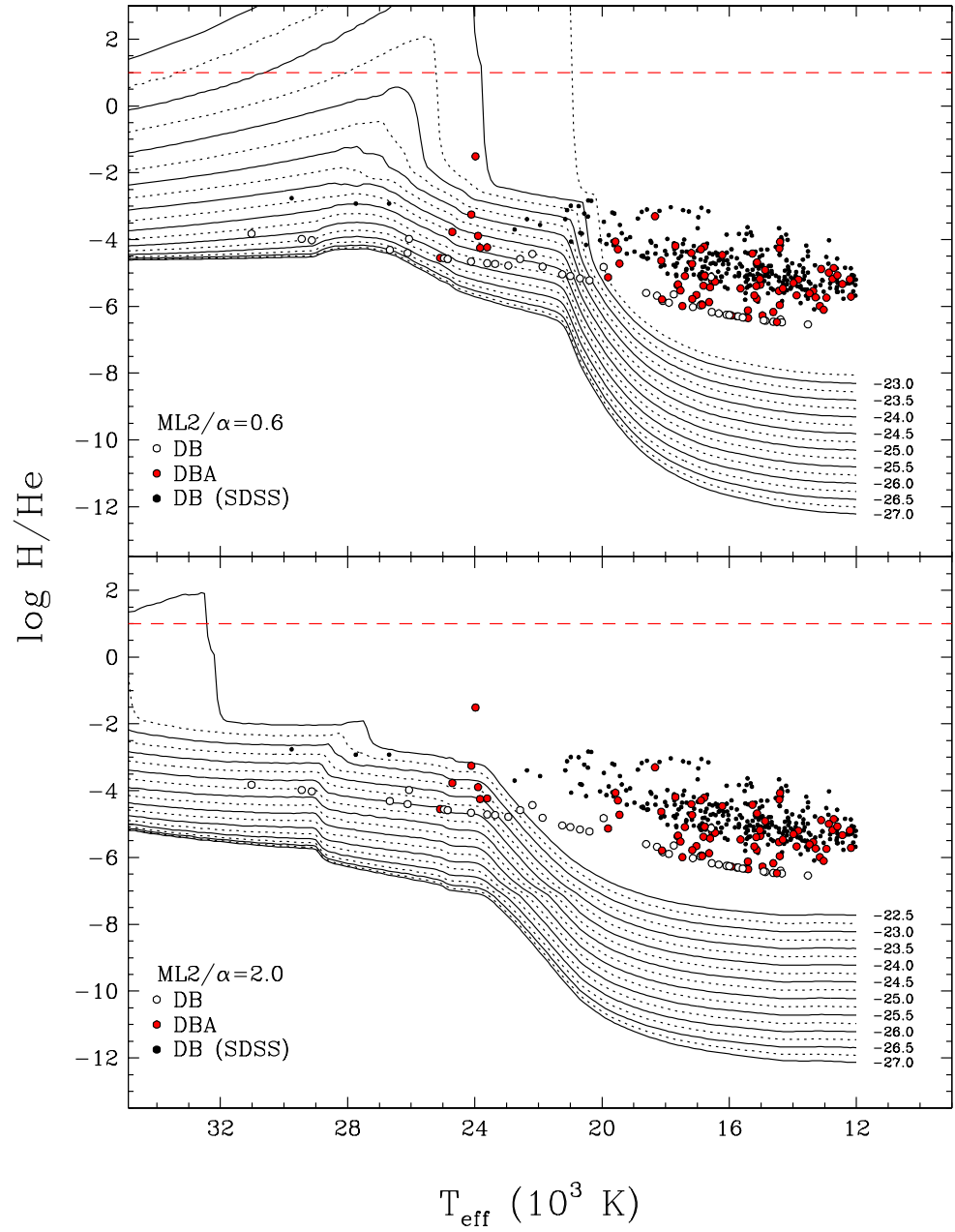}
\caption{Same as Figure \ref{HHe_cus_ML2}, but for a 0.6 \msun\ pure
  helium envelope white dwarf subject to accretion. Each curve is
  labeled with the corresponding accretion rate in solar mass per year
  (on a logarithmic scale).
\label{HHe_acc_ML2}}
\end{figure}

The photospheric hydrogen abundances as a function of $\Te$ predicted
by our convective dilution model in the presence of accretion are
displayed in Figure \ref{HHe_acc_ML2}. These results confirm our
previous expectations, that the convective dilution process will occur
in the presence of accretion, but only for extremely low accretion
rates of $\Dot{M}\lesssim 10^{-23}$ \msun\ yr$^{-1}$. This upper limit
is 2 to 5 orders of magnitude smaller than the accretion rate required
to account for the observed hydrogen abundances in DBA white dwarfs
(see Figure 15 of RBF18 in particular). But according to our
calculations, such high accretions rates will produce DA star
progenitors with hydrogen layers that are thick enough to prevent any
form of DA-to-DB transformation in the appropriate range of
temperature. Note that we exclude here the scenario where large bodies
such as small planets, comets, or disrupted asteroids have been
accreted {\it after mixing has occurred}, as often invoked in objects
such as GD 16, GD 17, GD 40, GD 61, G241-6, PG 1225$+$079, and HS
2253$+$8023, which all show evidence of accretion of water-rich
material (\citealt{Raddi15}, \citealt{fusillo17}). However, we do not
believe that this particular scenario applies to the bulk of DBA white
dwarfs, although it cannot be completely ruled out.

Hence, we conclude that the amount of hydrogen observed in DBA stars
cannot be explained as having a residual origin --- the leftovers from
the convective dilution scenario --- nor can it be accounted for by
accretion, at least in terms of an average accretion rate. In the next
section, we present an alternative possibility as a solution to the
problem of hydrogen in DBA white dwarfs.

\section{A CONVECTIVE DREDGE-UP MODEL}\label{dredge}

\subsection{The Hydrogen Diffusion Tail}

As part of an independent investigation aimed at studying the spectral
evolution of hot white dwarfs (A.~B\'edard et al.~2020, in
preparation), we performed time-dependent diffusion calculations (as
opposed to static models), with a particular interest in following the
diffusion of hydrogen towards the surface --- the so-called float-up
model --- the mechanism believed to be responsible for turning most
hot, helium-atmosphere white dwarfs into DA stars by the time they
reach the blue (hot) edge of the DB gap \citep{FW87}. These time-dependent
diffusion calculations are based on a completely new and modern
implementation of an evolutionary code named STELUM --- a short name
for tools for STELlar modeling from Universit\'e de Montr\'eal ---,
geared specifically toward stellar seismology and spectral/chemical
evolution \citep{brassard15}. A specificity of this code is that
evolutionary computations are producing complete models of stars: down
from the core up to, and including, the atmosphere.

As an illustrative example, we discuss here a particular white dwarf
evolutionary track at 0.6 \msun. We begin with a typical PG1159
progenitor that has evolved through late thermal flash episodes, where
the content of both hydrogen and helium has been significantly
depleted \citep{WH06}. In such a star, the stellar envelope has an
extremely uniform chemical composition through intense convective
mixing within the deep stellar envelope. This homogeneous chemical
profile is maintained by a strong stellar wind, which eventually dies
out at the end of the PG1159-phase \citep{quirion12}.  Our initial
stellar envelope is thus composed of a homogeneous mixture of H/He/C/O
all the way from the surface to a depth of $\log
M_{\rm{env}}/M_\star=-1.87$, surrounding an inner C/O core. The
chemical composition of the atmosphere and envelope of our initial
model is characterized by mass fractions of $X(\rm H)=10^{-6}$,
$X(\rm{He})=0.42$, $X(\rm C)=0.43$, and $X(\rm O)=0.15$, corresponding
to a total amount of hydrogen of $\log M_{\rm{H}}/M_\star=-7.62$ (or
$\log M_{\rm{H}}/M_\odot=-7.84$), thoroughly diluted within the deep
stellar envelope. Possible residual nuclear burning of hydrogen is
neglected.

\begin{figure}[bp]
\centering
\includegraphics[keepaspectratio=true,width=0.8\linewidth]{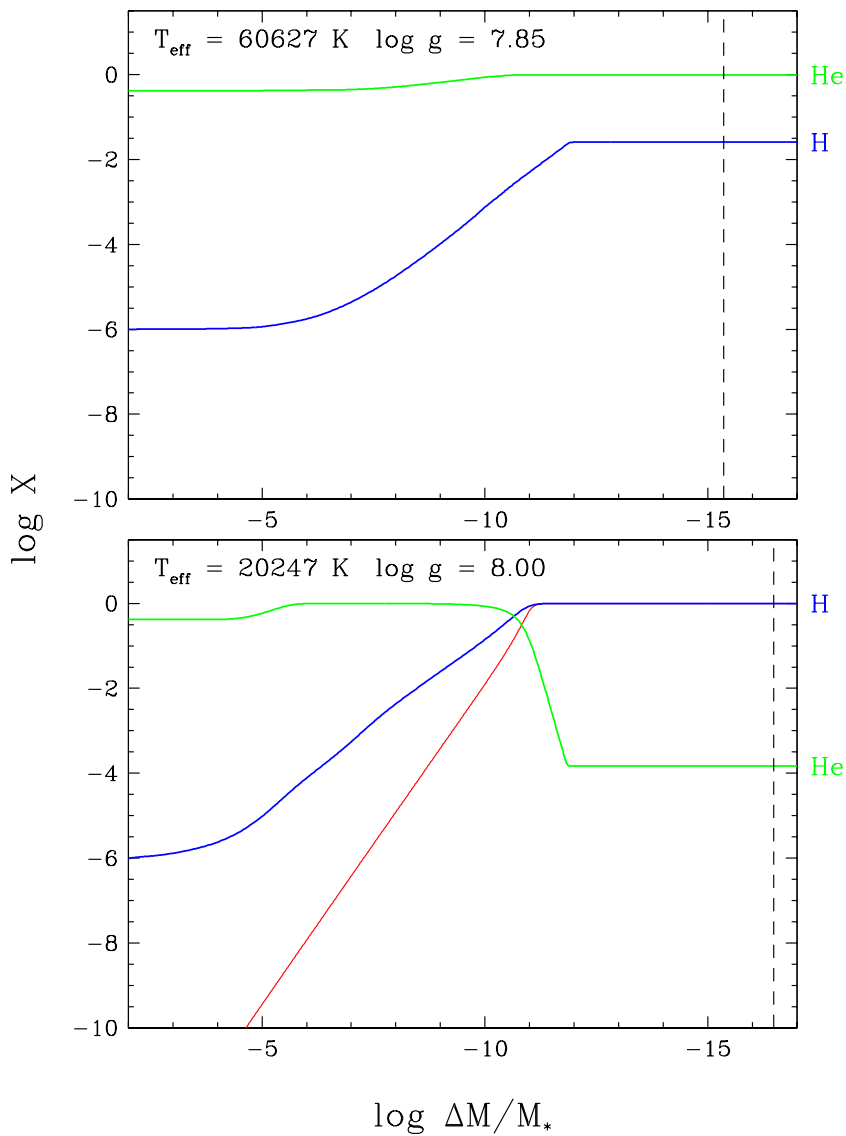}
  \caption{Snapshots at two different epochs of the hydrogen
      (blue) and helium (green) mass fractions as a function of depth,
      taken from time-dependent calculations of a 0.6 $M_\odot$ white
      dwarf with a total hydrogen mass fraction of $X({\rm
        H})=10^{-6}$ in the envelope. Also displayed in the bottom
    panel is a diffusive equilibrium profile containing
    $\sim$$10^{-11}~M_\star$ of hydrogen (red line). In both
      panels the location of the photosphere is indicated by the
      vertical dashed line.}\label{FigureRolland}
\end{figure}

As time proceeds, some helium rapidly rises to the surface,
transforming the PG1159 star into a DO white dwarf, and eventually
into a DB white dwarf at lower effective temperatures. Hydrogen also
diffuses upward, but much more slowly, eventually reaching the outer
surface, at which point the object becomes a DA white dwarf. Since
this is only an illustrative example, it is outside the scope of this
paper to discuss all the relevant details of this particular
evolutionary track. In particular, the convective dilution mechanism
has not been implemented yet. Instead, we show in Figure
\ref{FigureRolland} snapshots at two different epochs of the hydrogen
and helium mass fractions as a function of depth. Note that these are
still exploratory calculations, which are extremely time-consuming
(several weeks of CPU time for one sequence), and we have yet to study
the effect of several mechanisms, in particular convective
overshooting, as discussed for instance in
\citet{cukanovaite19}. Nevertheless, our models depict the important
features we want to address in this section.

In the top panel of Figure \ref{FigureRolland}, we show the abundance
profiles of hydrogen and helium at $\Te= 60,627$~K, at a point where
the white dwarf still appears as a DO star, with the photospheric
hydrogen abundance too low to be detected. The photospheric helium
abundance is already high, however, but this is just because the
initial helium content is already much larger than that of
hydrogen. More of interest in the present context are the abundance
profiles displayed in the bottom panel of Figure \ref{FigureRolland},
at a temperature of $\Te= 20,247$~K, where we expect the convective
dilution process to occur. Note that in this particular sequence, the
amount of hydrogen accumulated in the outer envelope, $\log M_{\rm
  H}/M_\star \sim -11$, is way too large for any dilution process to
occur, and this star will turn into a DC white dwarf --- or a
helium-rich DA star --- way below $\Te=10,000$~K. Nevertheless, what
our calculations reveal, is that even if hydrogen is a trace element
in the envelope, its abundance profile is always far from equilibrium
because of the extremely large diffusion timescales in the deeper
layers. In particular, near $\Te\sim 20,000$~K, only $\log M_{\rm
  H}/M_\star \sim -11$ of hydrogen has accumulated in the outer layers
of the white dwarf, which represents an extremely small fraction ---
about 0.04\% --- of the total hydrogen content present in the entire
star ($\log M_{\rm{H}}/M_\star=-7.62$). As a comparison, we show the
hydrogen profile (in red) in full diffusive equilibrium that has a
similar thickness of the outer hydrogen layer. Note that in their
calculations, \citet{MV91} assumed that this full diffusive
equilibrium has been reached at every $\Te$ value.

The most important aspect of these calculations is that substantial
amounts of hydrogen may still be located in the deeper layers of a
white dwarf, regardless of the amount already accumulated at the
surface. This is an unavoidable consequence due to the extremely long
diffusion timescales of hydrogen in the deeper layers; note that a
similar result has been obtained in the case of the helium diffusion
tail by \citet[][see their Figure 1]{dehner95}. Since the photosphere
is usually close to the surface at these effective temperatures
(see Figure \ref{FigureRolland}), the majority of the hydrogen in the
envelope would thus sit in a deep reservoir, inaccessible to
spectroscopic observations.

\subsection{Approximate Abundance Profiles}

We now attempt to take into account the conclusions of the previous
subsection --- summarized in Figure \ref{FigureRolland} --- into our
convective dilution simulations described in Section \ref{convdil}. To
this end, we thus consider that the overall hydrogen abundance profile
can be represented by the combination of a deep reservoir and a
surface contribution. Because of this massive reservoir, the
superficial hydrogen layer keeps building up as a function of time (or
decreasing effective temperature), but for the sake of simplicity, we
assume here that the only relevant abundance profile is that when
convective dilution occurs near $\Te\sim20,000$~K.

Due to the extremely large diffusion timescales found in the deeper
layers, the hydrogen reservoir remains virtually unaffected in the
earlier white dwarf phases, and only a small fraction of the total
hydrogen mass has reached the surface at the temperature where the
convective dilution process takes place (0.04\% in the example
above). We therefore assume that the hydrogen mass fraction is nearly
constant for $\log\Delta M/M_\star\lesssim -4$. In the outer layers,
however, the abundance profile will, by comparison, reach its
equilibrium value more quickly than in the deeper regions. For the
surface contribution, we thus rely on full diffusive equilibrium
profiles similar to those already described in Section \ref{modstruct}
(see also \citealt{MV91} and \citealt{manseau16}). Below this
superficial hydrogen layer of mass $M_{\rm H}$, we use two simple
power laws (with exponents $-9/10$ and $-3/4$) to describe the
hydrogen tail displayed in Figure \ref{FigureRolland}. In the deeper
regions, we then connect this tail to the massive reservoir at
$\log\Delta M/M_\star\sim -4$. This connection defines the hydrogen
mass fraction in the deeper envelope, and by definition the total
initial mass fraction of the star. Note that the exact value is not
important in the present context since the bottom of the H/He
convection zone never reaches these deeper regions.

\begin{figure}[bp]
\centering
\includegraphics[keepaspectratio=true,width=0.8\columnwidth]{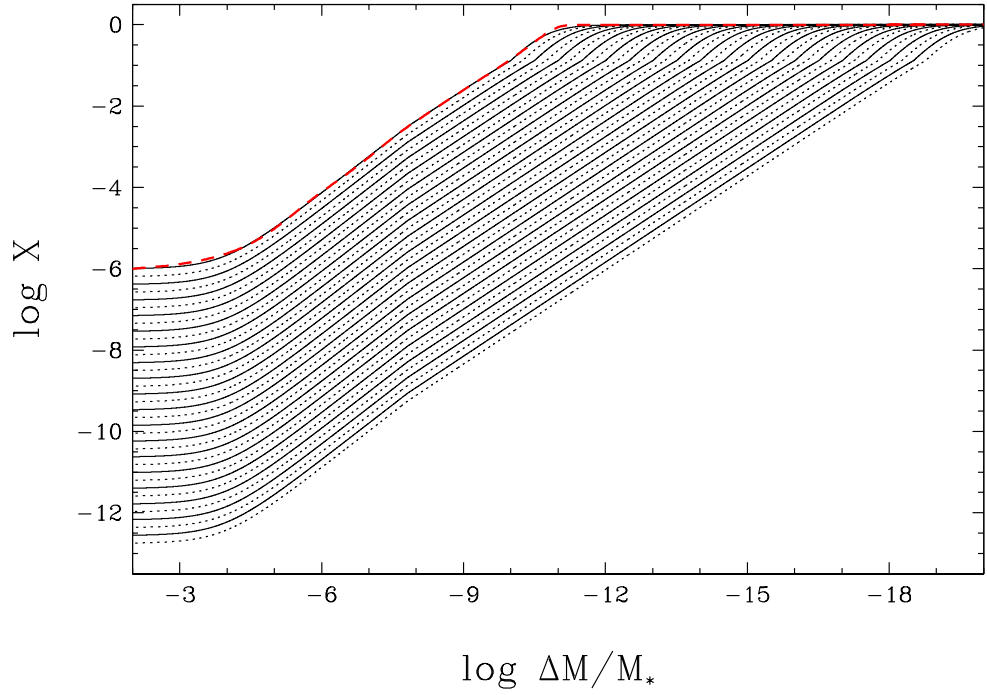}
\caption{Hydrogen mass fraction as function of depth for our
  approximate abundance profiles in a 0.6 \msun\ white dwarf. The
  thickness of the hydrogen surface layer ranges from $\log M_{\rm
    H}/M_\odot=-19.0$ to $-10.5$ by steps of 0.25 dex (from right to
  left). The detailed hydrogen profile, reproduced from Figure
  \ref{FigureRolland}, is represented by the red dashed line together
  with our profile for $\log M_{\rm
    H}/M_\odot=-10.35$. \label{res0_xmass}}
\end{figure}

Examples of these approximate hydrogen abundance profiles are
displayed in Figure \ref{res0_xmass} for various values of the
hydrogen surface layer mass, $M_{\rm H}$. For comparison, we also
reproduced the detailed profile from Figure \ref{FigureRolland}, which
is perfectly represented by our approximate profile with $\log M_{\rm
  H}/M_\odot=-10.35$. We can also see that with decreasing value of
$M_{\rm H}$, the hydrogen mass fraction in the deeper layers decreases
as well, as expected. We acknowledge that these profiles represent
only an approximation of the true profiles, but we are confident that
our approach represents a good characterization of the fully
time-dependent, evolutionary calculations, and that we can capture the
essential information required to revise the spectral evolution of DBA
white dwarfs.

\subsection{Convective Dredge-up Simulations}

In order to follow the spectral evolution of white dwarfs subject to
convective dredge-up, we first assume that all of the hydrogen
belonging to the surface contribution (as described above) has reached
the surface and obeys our new stratified envelope structures. With
this approximation, we only need to take into account the contribution
from the deep hydrogen reservoir. We then follow the same procedure
outlined in Section \ref{convdil}, but the target value of $M_{\rm H}$
is updated at every temperature step by adding the mass of hydrogen
dredged-up by the growing H/He convection zone. Since the reservoir
contains only trace amounts of hydrogen, the extent of the convection
zone will not be affected (see RBF18), and the transition temperatures
that take into account the convective dredge-up process are virtually
identical to those given in Table \ref{table_dilT}. In other words,
the convective dilution process depends on the amount of hydrogen
accumulated at the surface, but it is not affected by the presence of
the deep hydrogen reservoir. However, the resulting hydrogen
abundances at the photosphere, after mixing has occurred, should be
significantly different.  We now test this new paradigm by comparing
the predictions of our convective dredge-up simulations with the
photospheric H/He abundances measured in DB and DBA white dwarfs.

\begin{figure}[bp]
\centering
\includegraphics[keepaspectratio=true,width=0.8\columnwidth]{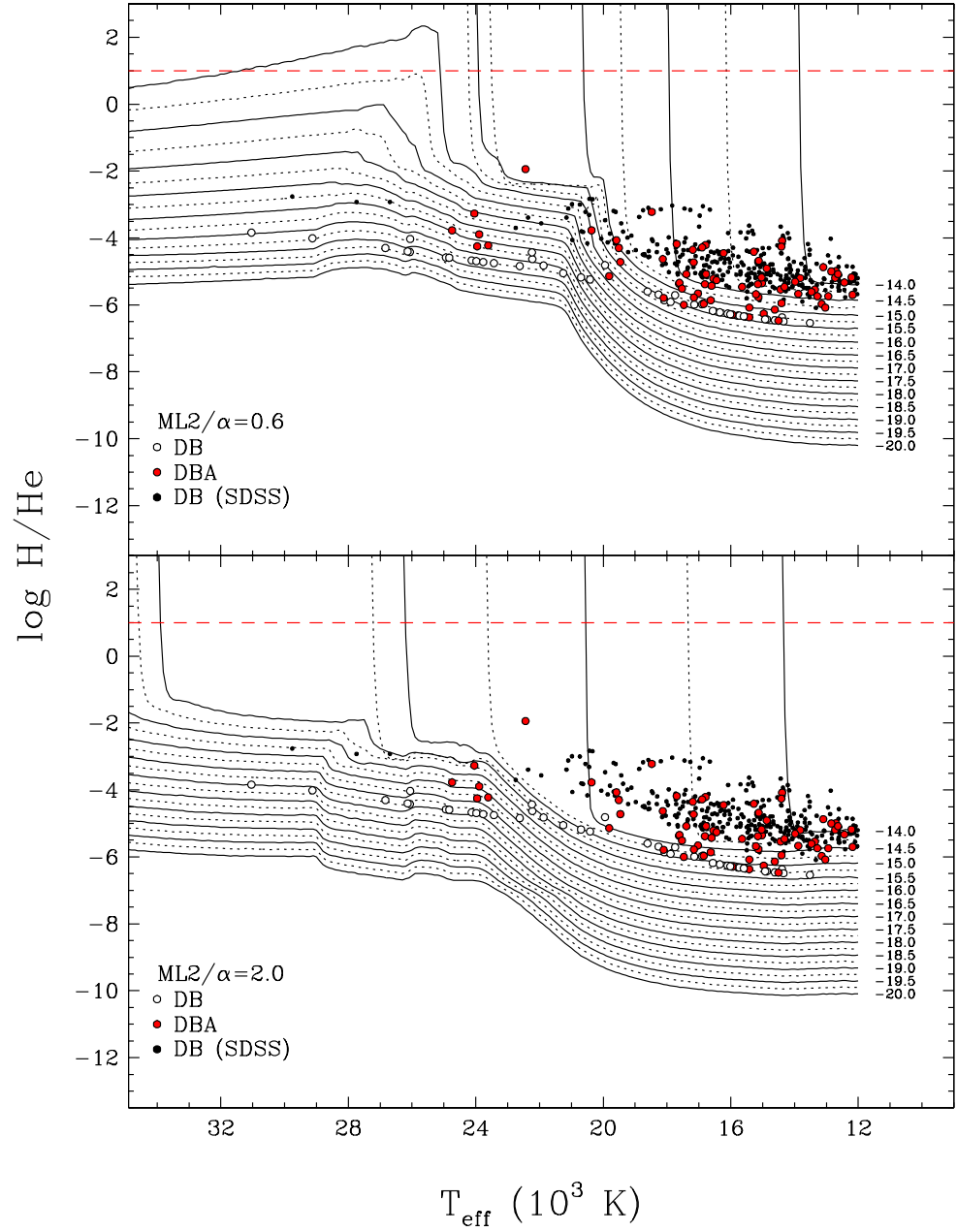}
\caption{Same as Figure \ref{HHe_cus_ML2}, but for a 0.6 \msun\ white
  dwarf undergoing convective dredge-up. Each curve is labeled with
  the corresponding initial hydrogen mass at the surface in solar mass
  (on a logarithmic scale).
\label{HHe_drg_ML2}}
\end{figure}

The results of our simulations are presented in Figure
\ref{HHe_drg_ML2}, which can be compared directly with Figure
\ref{HHe_cus_ML2}, where the dredge-up process has been ignored. Our
results clearly demonstrate that {\it the addition of the deep
  hydrogen reservoir represents the key element to reproduce the
  observed hydrogen abundances in the bulk of DBA white dwarfs}. The
H/He abundance ratio at $\Te=15,000$~K is predicted $\sim$2.3 dex
larger at the photosphere by including the dredge-up process. With
this scenario, a significant fraction of all DBA white dwarfs can be
explained with an initial superficial hydrogen layer mass lower than
$\log M_{\rm{H}}/M_\odot\sim -14$, combined with a progressive
enrichment from the deeper envelope layers. At the same time, the
hottest DBA stars can still be explained by surface layers thinner
than $\log M_{\rm{H}}/M_\odot\sim -16.5$, but the predicted hydrogen
abundances at lower temperatures would probably be undetectable, even
within the context of the dredge-up scenario. We note that some of the
DBA white dwarfs in Figure \ref{HHe_drg_ML2} would require even larger
hydrogen abundances, but we believe that such large abundances could
probably be produced by using more accurate hydrogen abundance
profiles than those displayed in Figure \ref{res0_xmass}, and more
importantly, by including some convective overshooting at the bottom
of the mixed H/He convection zone (see for instance the important work
by \citealt{Cunningham2019} and \citealt{cukanovaite19}).

\section{DISCUSSION}\label{conclu}

We presented a series of improved envelope models, both with
stratified and mixed chemical compositions, with the aim of simulating
the convective dilution process, where a thin radiative superficial
hydrogen layer is gradually eroded by the deeper helium convection
zone. This process is believed to be responsible for the DA-to-DB
transformation below the red edge of the DB gap near
$\Te\sim30,000$~K, although our calculations reveal that the
convective dilution mechanism becomes efficient only below
$\Te\sim20,000$~K, when the bottom of the helium convective envelope
plunges deep into the stellar interior. If the mass of the outer
hydrogen layer is too large ($\log M_{\rm H}/M_\odot\gtrsim -14$),
however, convective energy transport in the underlying helium envelope
is suppressed, thus preventing this convection dilution process to
occur. Even though the actual mixing process is obviously a
complicated, time-dependent dynamical process, we believe that our
static envelope models provide a fairly accurate description of the
DA-to-DB transformation, in particular in terms of the effective
temperatures at which this process occurs as a function of the
hydrogen layer mass, as given in Table \ref{table_dilT}.

We note that our transition temperatures differ slightly from those
provided in Table 1 of \citet{MV91}, probably because of significant
differences between their approach and ours, the most important of
which is the distribution of the total hydrogen mass within the
stellar envelope. In our calculations, hydrogen is distributed
entirely within the mixed H/He convection zone as well as at the
stellar surface where it is assumed to be in diffusive
equilibrium. More importantly, there is no hydrogen below the
convection zone (note that we refer here to our calculations of the
convective dilution process described in Section \ref{convsim}, where
there is no deep hydrogen reservoir). In MacDonald \& Vennes, there is
also a deep hydrogen diffusion tail underneath the mixed H/He
convection zone (see their Figure 6), which sometimes may involve most
of the total hydrogen content. In their calculations, the authors
indeed assume that a full equilibrium has been reached at every $\Te$
value, allowing any amount of hydrogen present in the star to achieve
a diffusive equilibrium, even below the convection zone. Hence at a
given effective temperature and hydrogen mass fraction in the
convection zone, the total integrated hydrogen mass in our
model is much smaller than in their model, thus explaining the
differences between the results in our Table \ref{table_dilT} and
their Table 1.

This being said, we believe our approach is probably more realistic in
the context of the convective dilution process. Let us assume for
instance that a hot DA white dwarf with a fully radiative hydrogen
layer of $\log M_{\rm H}/M_\odot=-15$ sits in diffusive equilibrium
on top of a helium envelope. When helium becomes convective, some of
the hydrogen layer will be mixed within the convection zone, and some
of it might leak at the bottom of the convective envelope, but
certainly not to the point of achieving a full diffusive equilibrium
profile. Hence it is probably better in the present context
to assume that there is no hydrogen below the convection zone, at
least when determining the DA-to-DB transition temperatures.

Although the exact amount of hydrogen present in the outer layers of
the DA progenitor determines the temperature at which the convective
dilution process will occur, the resulting hydrogen-to-helium
abundance ratio measured at the photosphere of DBA white dwarfs may
have little to do with this residual mass of hydrogen, in particular
if there is a massive reservoir of hydrogen present in the deep
stellar interior, which can be dredged-up to the surface by the mixed
H/He convective envelope. This is the new paradigm we explored
quantitatively in Section \ref{dredge}.

The overall spectral evolution picture that emerges from these
calculations is the following. As discussed above, we begin with a
typical PG1159 progenitor that has evolved through late thermal flash
episodes \citep{WH06}. In such a star, the content of both hydrogen
and helium has been significantly depleted, and the stellar envelope
has an extremely uniform chemical composition through intense
convective mixing within the deep stellar envelope. The exact amount
of hydrogen and helium thoroughly diluted within the deep stellar
envelope is unknown, but might be extremely small. During the earlier
phases, this homogeneous chemical profile is maintained by a strong
stellar wind, which eventually dies out at the end of the PG1159-phase
\citep{quirion12}. As the star cools off, helium begins to diffuse
upward, thus transforming the PG1159 progenitor into a DO white
dwarf. Eventually, hydrogen will also reach the surface, producing DAO
white dwarfs, which will in time turn into chemically stratified DAO
or DAB white dwarfs, and eventually into DA stars depending on the
amount of hydrogen that has accumulated at the surface (see, e.g.,
\citealt{manseau16}). The important point in this scenario is that the
amount of hydrogen that has reached the surface at a given point in
time --- i.e., at a given effective temperature --- may represent only
a small fraction of the total mass present in the stellar
envelope. Most importantly, the hydrogen abundance profile remains
always far from equilibrium because of the extremely large diffusion
timescales of hydrogen in the deeper layers.

If the amount of hydrogen accumulated at the surface of the white
dwarf is small enough --- $\log M_{\rm H}/M_\odot\lesssim-14$
according to our results given in Table \ref{table_dilT} --- the
convective dilution of this thin hydrogen layer by the more massive
underlying convective helium envelope will occur, at a mixing
temperature that depends on the hydrogen layer mass. At this point,
the bottom of the mixed H/He convective reaches deep into the stellar
interior, where large amounts of hydrogen may still reside. The
convective dredge-up process that follows may potentially carry large
amounts of hydrogen to the stellar photosphere, depending on the exact
quantity of hydrogen present in the deeper interior. In this scenario,
the amount of hydrogen originally present at the surface, prior to the
convective dilution process, represents only a negligible fraction of
the total mass of hydrogen present in the mixed H/He convective
envelope after the dredge-up process has occurred.

On the other hand, if the amount of hydrogen accumulated at the
surface of the white dwarf progenitor is too small --- $\log M_{\rm
  H}/M_\odot\lesssim-16$ according to \citet{manseau16} --- the star
will never become a genuine DA white dwarf, and will only appear as a
stratified DAB star, as discussed in Section \ref{convsim}. In this
context, we interpreted the hot DBA white dwarfs near
$\Te\sim24,000$~K in Figure \ref{HHe_cus_ML2} as descendants of such
hydrogen-poor progenitors. We also mentioned that with time, these DBA
white dwarfs would eventually turn into DB stars, with no detectable
traces of hydrogen, when the growth of the mixed H/He convective zone
below $\Te\sim 20,000$~K completely dilutes any amount of hydrogen
present in the outer layers. However, within the context of the
convective dilution scenario, it is also possible that the growth of
the convection zone is accompanied by a hydrogen enrichment from the
deeper layers into the photospheric regions, in which case the star
could remain a DBA white dwarf.

If the convective dilution scenario is the correct explanation for the
observed abundances of hydrogen in the bulk of DBA white dwarfs, we
are forced to conclude that the pure DB stars below $\Te=20,000$~K ---
those that show no H$\alpha$ absorption feature whatsoever --- must
have very little hydrogen left in their stellar envelope, even in the
deeper regions, most likely because of very intense and repeated late
thermal flash episodes in their earlier evolutionary phases. If this
is indeed the case, these hydrogen-free DB stars could represent the
progenitors of the cool DQ white dwarfs, in which hydrogen is rarely
detected spectroscopically, as opposed to other cool, non-DA
degenerates (the DZA stars in particular).

Finally, it is clear that accretion of hydrogen from external sources
--- mostly from comets, disrupted asteroids, and small planets ---
play an important role in the spectral evolution of helium-atmosphere
white dwarfs. In fact, the extreme hydrogen and metal abundances
measured in some of these objects can only be explained by such
accretion mechanisms. However, as discussed in Section \ref{results},
accretion needs to proceed {\it after} the convective dilution process
has occurred, otherwise, even modest average accretion rates would
build over time a hydrogen layer so thick that the DA-to-DB transition
becomes impossible. For these reasons, we believe that the convective
dredge-up model, proposed in this paper to explain the origin of
hydrogen in the {\it bulk of DBA white dwarfs}, represents a more
plausible scenario. One could even argue that, if the overall picture
described above is correct, hydrogen enrichment due to convective
dredge-up is an unavoidable outcome. Additional time-dependent
calculations such as those illustrated in Figure \ref{FigureRolland}
should eventually shed some light on this issue.

\acknowledgements We are grateful to C.~Genest-Beaulieu and A.~B\'edard
for useful discussions. This work was supported in part by the NSERC
Canada and by the Fund FRQ-NT (Qu\'ebec). G.F.~also acknowledges the
contribution of the Canada Research Chair Program.

\clearpage
\bibliography{ms}{}
\bibliographystyle{apj}

\end{document}